\documentclass[12pt]{article}
\usepackage{amssymb}
\usepackage{amsfonts}
\usepackage{amsmath}
\usepackage{amsthm}

\title{Has CHSH-inequality any relation to EPR-argument?}

\author{Andrei Khrennikov\\International Center for Mathematical Modeling \\
in Physics, Engineering, Economics, and Cognitive Science\\
Linnaeus University, V\"axj\"o, Sweden}

\begin{document}
\maketitle

\begin{abstract} 
We emphasize the role of the precise correlations loophole in attempting to  connect the CHSH type inequalities with the EPR-argument.
The possibility to test theories with hidden variables experimentally by using such inequalities is questioned. The role of the original Bell 
inequality is highlighted. The interpretation of the CHSH inequality in the spirit of Bohr, as a new test of incompatibility, is presented. 
The positions of Bohr, Einstein, Podolsky, Rosen, Bell, Clauser, Horne, Shimony,  Holt, and De Broglie are enlightened.
\end{abstract}

\section{Introduction}

The recent success in performing clean and loophole free experiments \cite{B1, B2, B3} testing violations of the Bell-type inequalities can make the impression that the long debate on
interpretation of violation of  these inequalities has been 
finally ended. Moreover, some authors, e.g., \cite{A1}, \cite{A2}, consider these experiments as the final accords in the long debate between Einstein and Bohr. Such authors  couple  
these inequalities with the Einstein-Podolsky-Rosen (EPR) framework \cite{EPR}, cf., however, \cite{KUP, AFTER}.
Nowadays it is widely claimed that {\it ``Bohr was right and Einstein was wrong''}. It is interesting that this formulation 
peacefully coexists with the statement that experiment confirms {\it ``quantum nonlocality.} 
The aim of this note is analyzing the After-Bell situation in quantum foundations, see also \cite{AFTER}. 

\section{Does CHSH inequality have any relation to the EPR framework?}

I claim that the answer to this question is negative.  The key point of the EPR framework is consideration of {\it precise correlations} and coupling them  with EPR elements of reality.
In particular, the EPR statement on quantum nonlocality (as an absurd alternative to incompleteness of QM) has meaning only this framework. Those few who read the original 
Bell's paper  \cite{B_EPR}, see also \cite{B},  know  that here Bell tried to mimic   the EPR framework \cite{EPR} by using hidden variables. However, the quantum counterpart of 
initial Bell's scheme was based on theoretical possibility of preparation of {\it singlet states}. At that time preparation of ensembles of high quality singlet states was totally impossible. 
 Bell understood well that his original inequality which was derived under assumption of perfect (anti-)correlations cannot be tested experimentally. 
And  he was happy to join Clauser, Horne, Shimony, and Holt  who used a new scheme  and derived CHSH-inequality \cite{CHSH}. It seems that in future Bell had never 
mentioned \cite{B} his original inequality derived in 1964, {\it the original Bell inequality.} The CHSH-scheme is not based on consideration of  precise (anti-)correlations. 
It provides the possibility  of performing experiment, even 
without clean technology for preparing singlet states.

Experimenting with the CHSH inequality \cite{CHSH} and inequalities based on the same scheme \cite{CHSH1, CHSH2, CHSH3, CHSH4} was extremely stimulating for development of quantum technologies.  It was also 
one of biggest challenges for experimenters in history of physics. Therefore it is impossible underestimate the value of the CHSH-type inequalities for physics. 
However, we have to be honest and say explicitly: 

\medskip

{\it The CHSH inequality and other inequalities which are not based on precise correlations have nothing to do with  the EPR framework and the Einstein- Bohr debate.} 
   
\medskip

Hence, statements as ``Closing the Door on Einstein and Bohr's Quantum Debate'', see \cite{A1}, are not justified.
To close this door, the original Bell inequality \cite{B_EPR} as to be tested. Nowadays the experimental technology is essentially more advance than in Bell's time.
In particular, very clean ensembles of singlet states can be prepared.  Photo-detectors of high efficiency  were already used in quantum experiments.  This
 makes the experiment on violation of the original Bell inequality at least less impossible than in Bell's time, see \cite{BO} for detailed analyzing the interplay between 
 detection efficiency  and the singlet state preparation. Obviously, such an experiment is even bigger challenge for experimental physics than the previous 
experiments on violations of inequalities based on the CHSH scheme.   
    
\section{Would be Bohr happy with nonlocal ``closing the door'' in his debate with Einstein?}
\label{BC}

From reading Bohr \cite{BR} and philosophers who put tremendous efforts to clarifying Bohr's views, e.g.,\cite{PL1}, we understand that for him quantum mechanics (QM) 
is a local theory. In particular, he did not explore the nonlocality alternative in his reply to Einstein \cite{BRR}.   
It is practically unknown that Bohr also had his own notion of an  element of reality known as  {\it phenomenon.} And this  is a local notion, see \cite{PLKHR}.
 
 Hence, the talks of people claiming   they are  Copenhagenists  
and at the same time  speaking about  quantum nonlocality are really misleading. They should honestly reject the Copenhagen interpretation  and 
explicitly say that not only Einstein, but also Bohr and other members of the Copenhagen school were wrong, because they were sure in quantum locality.

\section{Incompatibility or nonlocality?}

By Bohr  {\it the complementarity principle }  is the fundamental principle of QM. Experimentally this principle is expressed in existence of {\it incompatible observables.}
Such observables cannot be jointly measured. Historically Heisenberg's uncertainty relation for the position and momentum observables was Bohr's starting point. In his reply to 
the EPR argument \cite{BRR} Bohr emphasized the role of the complementarity principle. By him the EPR argument does not bring anything new to quantum foundations.
  He stressed  that the complementarity principle is about the position and momentum of  {\it a single particle.} 
And quantum interference experiments, as the two slit experiment,  demonstrate incompatiblity of these observables. Bohr did not find anything new in the EPR-argument. 
For him,  quantum interference is the basic mystery of quantum mechanics. There is no need in ``additional mysteries'' such as quantum nonlocality.

The CHSH equality and other Bell type inequalities which are not based on precise correlations are just additional statistical tests for the complementarity principle.
In the CHSH scheme,  there are considered four observables $A_i$ and $B_i, i=1,2,$ such that $[A_i, B_j]=0.$ Hence, pairs $A_i, B_j$ can be measured jointly  and the 
corresponding correlations $\langle A_i B_j\rangle$ can be calculated. 
They violate the CHSH inequality for some state if and only if     $[A_1, A_2]\not=0$ and $[B_1, B_2]\not=0,$ i.e., the $A$-observables as well as 
the $B$-observables are incompatible, cf. with Bohr's reply \cite{BRR} to EPR paper \cite{EPR}.  This incompatiblity is crucial in the CHSH framework. 

For Bohr, violation of the CHSH inequality is just a tricky form of expression of interference between projections of spin or polarization on different axes. 

\section{EPR-framework: a loophole from subquantum world to quantum experiment}

Bohr's reply to Einstein is often commented as unclear and misleading. And there is a point. We can wonder: How can complementarity help in explanation of the perfect
correlations?   In no way! But, for Bohr, there was no need in ``explanation'' of their origin. QM is an operational formalism concerning    measurement outputs. The formalism 
predicts the existence of the EPR-states. And, for Bohr, this is the final point of the scientific treatment of this problem. 

However, these correlations are intriguing and some people are seeking their explanation. Of course, such an explanation can be generated only in some subquantum theory. 
What is the key point of the EPR-argument? This is coupling of elements of such a hypothetical subquantum theory with measurement outputs, 
the elements of reality with outputs of measurements for the EPR-states. In principle, this coupling can be used as a loophole in the Copenhagen doctrine. 
One can try to test the predictions of hypothetical subquantum theories. But such tests are meaningful  only for the EPR-states as, e.g., the singlet state.       

\section{Can the CHSH-scheme be used to test experimentally hidden variable theories?}
 
I claim that the answer to this question is negative.  Here ``experimentally'' is the crucial word. As was repeatably pointed out, the CHSH-inequality is not about 
precise correlations. Therefore we cannot use the EPR loophole from the subquantum world to quantum experiment. {\it There is no reason to assume that subquantum 
correlations expressed mathematically in terms of hidden variables coincide with the experimental correlations predicted by QM.} The correlations given by integrals 
with respect to the distribution of hidden variables satisfy the CHSH-inequality, but generally they have no relation to correlations obtained in experiment. In fact, this was 
De Broglie's viewpoint, see  \cite{De Broglie, AFTER}. This viewpoint match the Bild conception of scientific theory which elaborated by Hertz and Boltzmann, see \cite{Hertz}.   

\section{Concluding remarks}

The main impact of experimental testing for CHSH-like inequalities is demonstration that correlations predicted by QM  can be preserved for very long distances.
The only foundational impact of such tests is the  confirmation of Bohr's complementarity principle. Such inequalities and tests on their violation cannot be 
used for testing hypothetical subquantum theories with hidden variables. It seems that only the original Bell inequality can be used for such a purpose.   
Until a loophole free test for the latter will be performed, the statements as ``Bohr was right and Einstein was wrong'' or     ``Closing the door on Einstein and Bohr's quantum debate''
are  not justified.


\begin{thebibliography}{99}
\bibitem{B1}   B. Hensen et al, {\it Experimental loophole-free violation of a Bell inequality using entangled electron spins separated by 1.3 km}, 
Nature 526, 682 (2015);  arXiv:1508.05949 [quant-ph] (2015)
\bibitem{B2} M. Giustina et al, A significant-loophole-free test of Bell's theorem with entangled photons, Phys. Rev. Lett. 115, 250401 (2015) 
[arXiv: quant-ph1511.03190].
\bibitem{B3} L. K. Shalm et al,  A strong loophole-free test of local realism, Phys. Rev. Lett. 115, 250402 (2015)  [arXiv:quant-ph1511.03189].
\bibitem{A1} Aspect A.,  Closing the door on Einstein and Bohr's quantum debate, Physics 8, 123,  2015
\bibitem{A2}  Wiseman H.,   Quantum physics: Death by experiment for local realism, Nature,  526, 649–650, 2015
\bibitem{EPR} Einstein, A., Podolsky, B. Rosen, N.: Can quantum-mechanical description of physical reality be considered complete?.
Phys. Rev. 47 (10), 777-780 (1935).
\bibitem{KUP} Kupczynski M.,  Can Einstein with Bohr debate on quantum mechanics be closed?  {\it Phil. Trans. Royal Soc.} A, {\bf 375}, 
N 2106, 2016039 (2017);  arXiv:1603.00266 [quant-ph].
\bibitem{AFTER} A. Khrennikov, After Bell.  {\it Fortschritte der Physik (Progress in Physics)} {\bf 65},  N 6-8, 1600014  (2017).
\bibitem{B_EPR} Bell, J.: On the Einstein-Podolsky-Rosen paradox. Physics, 1, 195-200 (1964).
\bibitem{B} Bell, J.: Speakable and Unspeakable in Quantum Mechanics. Cambridge Univ. Pres, Cambridge (1987).
\bibitem{CHSH} Clauser J F, Horne M A, Shimony A and Holt R A  1969 Proposed experiment to test local hidden-variable theories.  
{\it Phys. Rev. Lett.} {\bf 23} (15)  880--884
\bibitem{CHSH1} Clauser J F and Horne M A 1974 Experimental consequences of objective local theories. {\it Phys. Rev.} D {\bf 10} 526-535.
\bibitem{CHSH2} Clauser J F and Shimony A 1978 Bell's theorem. Experimental tests and implications. {\it Rep. Prog. Phys.} {\bf 41} 1881-1927. 
\bibitem{CHSH3} Eberhard Ph H 1993 Background level and counter efficiencies required for a loophole-free Einstein-Podolsky-Rosen experiment.
 {\it Phys. Rev. A}, {bf 47},   477-750.
\bibitem{CHSH4} A. Khrennikov,  S. Ramelow,  R. Ursin, B. Wittmann, J. Kofler, I.  Basieva,
On the equivalence of the Clauser-Horne and Eberhard inequality based tests. {\it Physica Scripta}  {\bf T163} 014019  (2014)
\bibitem{BO} A. Khrennikov and I. Basieva, Towards experiments to test violation of the original Bell inequality.
{\it Entropy}, 20(4), 280  (2018); https://doi.org/10.3390/e20040280.
\bibitem{BR} Bohr N.,   The philosophical writings of Niels Bohr, 3 vols.,  Woodbridge, Conn., Ox Bow Press, 1987
\bibitem{PL1} Plotnitsky A.,  Reading Bohr: Physics and philosophy, Springer, Dordrecht, 2006
\bibitem{BRR} Bohr N., Can quantum-mechanical description of physical reality be considered complete? Phys. Rev., 48, 696-702, 1935
\bibitem{PLKHR} A. Plotnitsky and A. Khrennikov, Reality without realism: On the ontological and epistemological architecture of 
quantum mechanics {\it Found. Phys.} {\bf  45}, N 10, 1269-1300 (2015).	
\bibitem{De Broglie}  L. De Broglie, The current interpretation of wave mechanics: a critical study. Elsevier, 1964.
\bibitem{Hertz} A. Khrennikov,	Hertz's viewpoint on quantum theory. 	arXiv:1807.06409 [quant-ph]

\end{thebibliography}
\end{document}